\documentclass[aps,preprint,nofootinbib]{revtex4}
\usepackage{amsmath}
\usepackage{graphicx}
\usepackage{color}
\usepackage{slashed}
\usepackage{subfigure}
\usepackage{tabularx}
\usepackage{booktabs}
\usepackage{mathtools}
\usepackage{multirow}
\usepackage{makecell}
\allowdisplaybreaks

\begin{document}

\title{Higgs boson decays to a fermion pair and a polarized $Z$ boson at NLO accuracy\\[0.7cm]}

\author{\vspace{1cm} Zi-Qiang Chen$^{1}$, Qi-Ming Feng$^2$, and Cong-Feng Qiao$^2$\footnote[1]{qiaocf@ucas.ac.cn, corresponding author} \\}

\affiliation{$^1$School of Physics and Materials Science, Guangzhou University, Guangzhou 510006, China\\
$^2$ School of Physics, University of Chinese Academy of Sciences, Beijing 100049, China
\vspace{2.6cm}}

\begin{abstract}
\vspace{0.5cm}
In this paper, we investigate the Higgs boson decay process $H\to f\bar{f}Z\to f\bar{f}\mu^-\mu^+$, where $f$ denotes any light fermions, at the next-to-leading order (NLO) accuracy.
The calculation is performed within the framework of single-pole approximation, in which the contributions of different polarization states of the $Z$ boson are considered separately.
Numerical results show that, by taking appropriate cut, the non-resonant background is negligible, and the NLO corrections are $1\%$--$4\%$ of the LO contributions.
We also find that the inclusion of NLO corrections can greatly reduce the dependence on the electroweak coupling scheme, and enhance the prediction reliability. 
We formulate a method for the extraction of the pseudo-observable $\Gamma^{H\to ffZ}$ from the $f\bar{f}\mu^-\mu^+$ signal.
\end{abstract}
\maketitle

\newpage

\section{Introduction}
The Higgs mechanism is an essential ingredient of the Standard Model (SM) of particle physics, as it explains the origin of mass of fundamental particles  \cite{Higgs:1964pj,Englert:1964et,Guralnik:1964eu}.
The observation of the Higgs boson in 2012 \cite{Aad:2012tfa,Chatrchyan:2012ufa} marked a key milestone in particle physics.
Further measurements of its spin, parity and couplings have shown no significant deviation from the SM predictions.
Nevertheless, it is conceivable that much more detailed and precise investigations are required to verify the mechanism of spontaneous symmetry breaking and search for new physics beyond the SM.

The decay of the Higgs boson into four leptons is one of the most important decay channels for the measurement of Higgs boson properties, as it has a clear and clean signature. 
To date, this channel has been extensively studied by the ATLAS and CMS Collaborations \cite{Chatrchyan:2013mxa,CMS:2014quz,ATLAS:2014kct,CMS:2014nkk,CMS:2015chx,CMS:2017dib,CMS:2017len,ATLAS:2017qey,ATLAS:2017azn,ATLAS:2018jym,Sirunyan:2019twz,ATLAS:2020rej,ATLAS:2020wny,CMS:2021ugl,CMS:2021nnc,ATLAS:2022qef,CMS:2023gjz}, with the purpose of determining the Higgs boson mass, width, spin-parity, as well its coupling strengths to other particles.
Within the framework of the SM, theoretical predictions for the partial decay widths have been calculated up to next-to-leading order (NLO) in electroweak (EW) coupling expansion \cite{Kniehl:1993ay,Bredenstein:2006rh,CarloniCalame:2006vr,Bredenstein:2006ha}, and have been matched to quantum electrodynamics parton-showers \cite{Boselli:2015aha}.
The analysis with higher-dimensional operators are performed in Refs. \cite{Artoisenet:2013puc,Boselli:2017pef,Brivio:2019myy},
and those based on beyond-the-SM theories with modified Higgs sectors can be found in Refs. \cite{Altenkamp:2018bcs,Altenkamp:2018hrq,Kanemura:2019kjg}.

\begin{figure}
\includegraphics[width=0.35\textwidth]{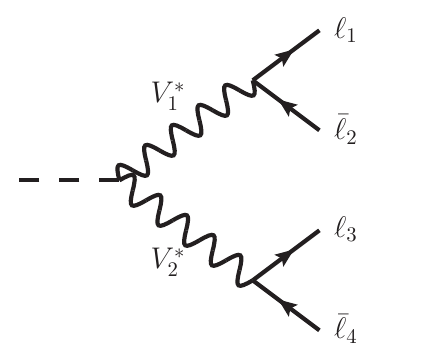}
\caption{Typical tree-level Feynman diagram for $H\to 4\ell$ process.}
\label{fig_FeytreeVV}
\end{figure}

When neglecting the Yukawa coupling between Higgs and leptons, the leading order (LO) process of Higgs to four leptons is mediated by two weak gauge bosons, as shown in Fig.\ref{fig_FeytreeVV}.
The corresponding Feynman amplitude can be factorized as:
\begin{equation}
\mathcal{M}^{H\to 4\ell}_{\rm Born}\sim \mathcal{M}^{H\to V_1^*V_2^*}_{\rm Born}\otimes \mathcal{M}^{V_1^{*}\to \ell_1\bar{\ell}_2 }_{\rm Born}\otimes \mathcal{M}^{V_2^{*}\to \ell_3\bar{\ell}_4}_{\rm Born},
\label{eq_facab}
\end{equation}
where $V_1$ and $V_2$ denote the intermediate weak gauge bosons.
By analyzing the kinematic distributions of the final state leptons, the polarizations of $V_1$ and $V_2$ can be reconstructed, which may give hints to new physics effects in the coupling of Higgs to polarized bosons \cite{Bruni:2019xwu}.
Theoretical studies on this topic can be found in Refs. \cite{Maina:2020rgd,Maina:2021xpe}, subject to the factorizable nature of LO amplitude. 
However, it can not be simply extended to the NLO due to the presence of non-factorizable corrections. This problem can be partly addressed by using the pole approximation technique \cite{Stuart:1991xk,Stuart:1991cc,Aeppli:1993rs}, where only those terms enhanced by the resonant of weak boson in propagator are taken into account. 
Since the Higgs mass is below the weak boson pair threshold, at most one weak boson can be on its mass shell.
Hence, in this situation the single-pole approximation (SPA) applies, which embodies the dominant contribution, rather than bothering to exploit the standard double-pole approximation \cite{Denner:2000bj,Baglio:2018rcu,Denner:2021csi}.
In the SPA, the amplitude is factorized as
\begin{equation}
\mathcal{M}^{H\to 4\ell}\overset{\rm SPA}{\sim} \mathcal{M}^{H\to \ell_1\bar{\ell}_2 V}\otimes \mathcal{M}^{V\to \ell_3\bar{\ell}_4},
\label{eq_facab}
\end{equation}
where $V$ denotes the nearly on-shell weak gauge boson.

The possibility of extending the LO polarization study to higher orders in Higgs decay was concerned in Ref. \cite{Maina:2021xpe},
while in practice no explicit calculation beyond the LO has been carried out so far.
For this sake, we calculate the NLO radiative corrections to the $H\to f\bar{f}Z\to f\bar{f}\mu^-\mu^+$ process, where $f$ denotes any light fermions, in the frame work of SPA.
To investigate the polarization effects of the intermediate $Z$ boson, the contributions of different polarization states are differentiated.
In addition to the SPA, we also introduce an on-shell approximation to the phase-space integral, which simplifies the calculation and facilitates the extraction of the pseudo-observable $\Gamma^{H\to ffZ}$.

The rest of the paper is organized as follows.
In Section II, we present the primary formulations employed in the calculation.
In Section III, we elucidate some technical details for the NLO calculation.
The numerical results are presented in Section IV.
The last section is remained for a summary.

\section{Formulations}
Let us start with the four-fermion decay process $H(p_1)\to f(p_2)+\bar{f}(p_3)+\mu^-(p_4)+\mu^+(p_5)$, whose decay width is given by 
\begin{equation}
\Gamma=\frac{1}{2m_H}\int \text{d}I_4(p_1;p_2,p_3,p_4,p_5) \; \left| \mathcal{M}^{H\to ff\mu\mu}(p_1;p_2,p_3,p_4,p_5) \right|^2,
\label{eq_4fwidth}
\end{equation}
where $m_H$ is the mass of Higgs boson, $\text{d}I_4$ stands for the four-body phase-space integral.
The summation notation over the final fermion spin and color states is dropped for brevity.
The explicit definition of $\text{d}I_4$ is
\begin{equation}
\int {\rm d}I_4(p_1;p_2,p_3,p_4,p_5)=\int \prod_{i=2}^5\bigg[ \frac{{\rm}d p_i^4}{(2\pi)^3}\delta(p_i^2)\bigg] (2\pi)^4\delta^4\bigg(p_1-\sum_{i=2}^5p_i\bigg).
\label{eq_I4}
\end{equation}
Inserting $1=\int {\rm d}p_{45}^4\; \delta^4(p_{45}-p_4-p_5)\int {\rm d}s_{45}\;\delta(s_{45}-p_{45}^2)$ into the right hand of Eq. (\ref{eq_I4}), the four-body phase-space integral can be decomposed as
\begin{equation}
\int {\rm d}I_4(p_1;p_2,p_3,p_4,p_5)=\frac{1}{2\pi}\int {\rm d}s_{45} \int {\rm d}I_3(p_1;p_2,p_3,p_{45})\int{\rm d}I_2(p_{45};p_4,p_5),
\label{eq_I4dcom}
\end{equation}
where $p_{45}$ is the momentum of the dimuon system, $s_{45}=p_{45}^2$, $\text{d}I_3$ and $\text{d}I_2$ stand for the three- and two-body phase-space integrals respectively.

 Equation (\ref{eq_I4dcom}) indicates that the decay width (\ref{eq_4fwidth}) can be written as $\Gamma=\int {\rm d}s_{45} f(s_{45})$, where $f(s_{45})\sim\int \text{d}I_3 \int \text{d}I_2 |\mathcal{M}|^2$.
For the case where the intermediate $Z$ boson nearly on-shell, we may perform a Laurent expansion of $f(s_{45})$ around $s_{45}=m_Z^2$.
The first term of the Laurent series can be constructed by using the SPA. That is\footnote{Here only the contributions of factorizable diagrams are taken into account.  The contributions of non-factorizable diagrams are expected to be very small according to Refs. \cite{Beenakker:1997bp,Beenakker:1997ir,Denner:1997ia,Denner:1998rh}.}
\begin{align}
\Gamma^\text{SPA}=&\frac{1}{2m_H}\frac{1}{2\pi}\int_{s_{45}^-}^{s_{45}^+} {\rm d}s_{45}\frac{1}{(s_{45}-m_Z^2)^2+m_Z^2\Gamma_Z^2}\int {\rm d}\hat{I}_3(p_1;p_2,p_3,\hat{p}_{45})\int {\rm d}\hat{I}_2(\hat{p}_{45};p_4,p_5)\nonumber \\
& \times \left| \mathcal{M}^{H\to ffZ}_\mu(p_1;p_2,p_3,\hat{p}_{45})\; g^{\mu\nu}\;\mathcal{M}_\nu^{Z\to \mu\mu}(\hat{p}_{45};p_4,p_5)\right|^2\nonumber \\
=&\frac{1}{\pi}\;{\rm arctan}\bigg(\frac{s_{45}-m_Z^2}{\Gamma_Z m_Z}\bigg)\bigg|^{s_{45}=s_{45}^+}_{s_{45}=s_{45}^-}\; \frac{1}{4m_H m_Z \Gamma_Z}\int {\rm d}\hat{I}_3(p_1;p_2,p_3,\hat{p}_{45})\int {\rm d}\hat{I}_2(\hat{p}_{45};p_4,p_5)\nonumber \\
&\times \left| \mathcal{M}^{H\to ffZ}_\mu(p_1;p_2,p_3,\hat{p}_{45})\; g^{\mu\nu}\;\mathcal{M}_\nu^{Z\to \mu\mu}(\hat{p}_{45};p_4,p_5)\right|^2.
\label{eq_narrwidth}
\end{align}
Here, ${\rm d}\hat{I}_3$ and ${\rm d}\hat{I}_2$ stands for the phase-space integrals with $\hat{p}_{45}^2=m_Z^2$;
$s_{45}^\pm$ denotes the upper and lower cuts on $s_{45}$.
Replacing the metric tensor with the polarization vectors, i.e. $g^{\mu\nu}\to -\sum_{\lambda}( \varepsilon^{* \mu}_{\lambda}\varepsilon^\nu_{\lambda}-\hat{p}_{45}^\mu \hat{p}_{45}^\nu/M_Z^2)$, the decay width can be written as\footnote{Since we neglect the lepton mass, the gauge term $\hat{p}_{45}^\mu \hat{p}_{45}^\nu/m_Z^2$ can be dropped off.}
\begin{equation}
\label{eq_wdcom}
\Gamma^\text{SPA}=\frac{1}{\pi}\;{\rm arctan}\bigg(\frac{s_{45}-m_Z^2}{\Gamma_Z m_Z}\bigg)\bigg|^{s_{45}=s_{45}^+}_{s_{45}=s_{45}^-}\;\sum_{\lambda=\pm,0}\sum_{\lambda^\prime=\pm,0} \Gamma^{H\to ffZ}_{\lambda\lambda^\prime} \frac{\Gamma^{Z\to \mu\mu}_{\lambda\lambda^\prime}}{\Gamma_Z},
\end{equation}
where
\begin{align}
&\Gamma^{H\to ffZ}_{\lambda\lambda^\prime}=\frac{1}{2m_H}\int {\rm d}\hat{I}_3(p_1;p_2,p_3,\hat{p}_{45})\big[\mathcal{M}^{H\to ffZ}_{\mu} \varepsilon^{*\mu}_{\lambda}\big]\big[\mathcal{M}^{H\to ffZ}_{\nu} \varepsilon^{*\nu}_{\lambda^\prime}\big]^*,\\
&\Gamma^{Z\to \mu\mu}_{\lambda\lambda^\prime}=\frac{1}{2m_Z}\int {\rm d}\hat{I}_2(\hat{p}_{45};p_4,p_5)\big[\mathcal{M}^{Z\to \mu\mu}_{\mu} \varepsilon^{\mu}_{\lambda}\big]\big[\mathcal{M}^{Z\to \mu\mu}_{\nu} \varepsilon^{\nu}_{\lambda^\prime}\big]^*.
\end{align}
Here, the momentum dependences of the amplitudes are suppressed for brevity. 

To describe the kinematic of the $Z\to \mu^-\mu^+$ process, we introduce $\theta^*$ and  $\phi^*$, the polar and azimuthal angles of $\mu^+$, in the rest frame of the $Z$ boson, with respect to the $Z$ flight direction in the Higgs rest frame.
For the definition of the polarization vector $\varepsilon_\lambda$, we take the expression presented in Ref. \cite{Maina:2020rgd}.
At the LO, it is known that the interference decay widths $\Gamma^{Z\to \mu\mu}_{\lambda\lambda^\prime}$ ($\lambda\ne \lambda^\prime$) vanish when the squared amplitude is integrated over the full range of the angle $\phi^*$ \cite{Maina:2020rgd,Denner:2021csi}:
\begin{equation}
\frac{\text{d}\Gamma^{Z\to \mu\mu}_{\lambda\lambda^\prime}}{\text{dcos}\theta^*}=\frac{1}{2m_Z}\frac{1}{32\pi^2}\int^{2\pi}_0 {\rm d}\phi^* \big[\mathcal{M}^{Z\to \mu\mu}_{\mu} \varepsilon^{\mu}_{\lambda}\big]\big[\mathcal{M}^{Z\to \mu\mu}_{\nu} \varepsilon^{\nu}_{\lambda^\prime}\big]^*
= 0,\; \text{for}\; \lambda\neq \lambda^\prime.
\label{eq_intfis0}
\end{equation}
We notice that this relation still hold at the NLO accuracy (see Appendix for a proof).
Hence the interference terms of Eq. (\ref{eq_wdcom}) can be dropped
\begin{align}
\frac{\text{d}\Gamma^\text{SPA}}{\text{dcos}\theta^*}=&\frac{1}{\pi}\;{\rm arctan}\bigg(\frac{s_{45}-m_Z^2}{\Gamma_Z m_Z}\bigg)\bigg|^{s_{45}=s_{45}^+}_{s_{45}=s_{45}^-}\nonumber \\
&\times \frac{1}{\Gamma_Z}\bigg(\Gamma^{H\to ffZ}_\text{T} \frac{\text{d}\Gamma^{Z\to \mu\mu}_\text{T}}{\text{dcos}\theta^*}+\Gamma^{H\to ffZ}_\text{L} \frac{\text{d}\Gamma^{Z\to \mu\mu}_\text{L}}{\text{dcos}\theta^*}+\Gamma^{H\to ffZ}_\text{A} \frac{\text{d}\Gamma^{Z\to \mu\mu}_\text{A}}{\text{dcos}\theta^*}\bigg).
\label{eq_mainformu}
\end{align}
Here, for convenience, we introduce the transverse (T),  longitudinal (L), and asymmetry (A) widths:
\begin{align}
\Gamma^{H\to ffZ}_\text{T}=\Gamma^{H\to ffZ}_{++}+\Gamma^{H\to ffZ}_{--},\quad &\Gamma^{Z\to \mu\mu}_\text{T}=\big(\Gamma^{Z\to \mu\mu}_{++}+\Gamma^{Z\to \mu\mu}_{--}\big)/2;\\
\Gamma^{H\to ffZ}_\text{L}=\Gamma^{H\to ffZ}_{00},\quad &\Gamma^{Z\to \mu\mu}_\text{L}=\Gamma^{Z\to \mu\mu}_{00};\\
\Gamma^{H\to ffZ}_\text{A}=\Gamma^{H\to ffZ}_{++}-\Gamma^{H\to ffZ}_{--},\quad &\Gamma^{Z\to \mu\mu}_\text{A}=\big(\Gamma^{Z\to \mu\mu}_{++}-\Gamma^{Z\to \mu\mu}_{--}\big)/2.
\end{align}
Note, equation (\ref{eq_mainformu}) is similar to the narrow width approximation, but with a overall factor $\frac{1}{\pi}\text{arctan}(\frac{s_{45}-m_Z^2}{\Gamma_Z m_Z})|^{s_{45}=s_{45}^+}_{s_{45}=s_{45}^-}$ to estimate the finite width effect of the nearly-on-shell $Z$ boson.
Taking $\text{d}\Gamma^{Z\to \mu\mu}_\lambda/\text{dcos}\theta^*$ as input, the pseudo-observable $\Gamma^{H\to ffZ}_\lambda$ can be extracted by confronting Eq. (\ref{eq_mainformu}) to the experiment measurement.

Finally, we have the following remarks on the treatment of the phase-space integral.
In the traditional pole approximation technique \cite{Denner:2000bj,Baglio:2018rcu,Denner:2021csi}, the full four-body phase space is used, and an appropriate momentum projection is introduced to keep the gauge invariance.
While in our case, the phase-space decomposition technique is used, and an on-shell approximation is applied to the sub-phase space, that is
\begin{align}
\int \text{d}I_3(p_1;p_2,p_3,p_{45})&=\frac{1}{(2\pi)^5}\int_0^{(m_H-\sqrt{s_{45}})^2} \text{d}s_{23}\; \frac{\lambda^{1/2}(m_H^2,s_{23},s_{45})}{64m_H^2}\int \text{d}\Omega_{45}^*\int \text{d}\Omega_2^* \nonumber \\
&\approx \frac{1}{(2\pi)^5}\int_0^{(m_H-m_Z)^2} \text{d}s_{23}\; \frac{\lambda^{1/2}(m_H^2,s_{23},m_Z^2)}{64m_H^2}\int \text{d}\Omega_{45}^*\int \text{d}\Omega_2^*\nonumber \\
& =\int \text{d}\hat{I}_3(p_1;p_2,p_3,\hat{p}_{45}),
\label{eq_dI3approx}
\end{align}
where $s_{23}=(p_2+p_3)^2$, $\lambda(x,y,z)=x^2+y^2+z^2-2xy-2xz-2yz$.
This treatment enables the separation of the $H\to f\bar{f}Z$ and $Z\to \mu^+\mu^-$ processes at the decay width level, which facilitates the extraction of the pseudo-observable $\Gamma^{H\to ffZ}_\lambda$ from the experiment data.
To reduce the theoretical error induced by the phase-space approximation, one may introduce a kinematic cut $s_{23}<\big(m_H-\sqrt{s_{45}^+}\big)^2$.

\section{Some details in the calculation}
\begin{figure}
\centering
\subfigure[]{
\includegraphics[scale=0.6]{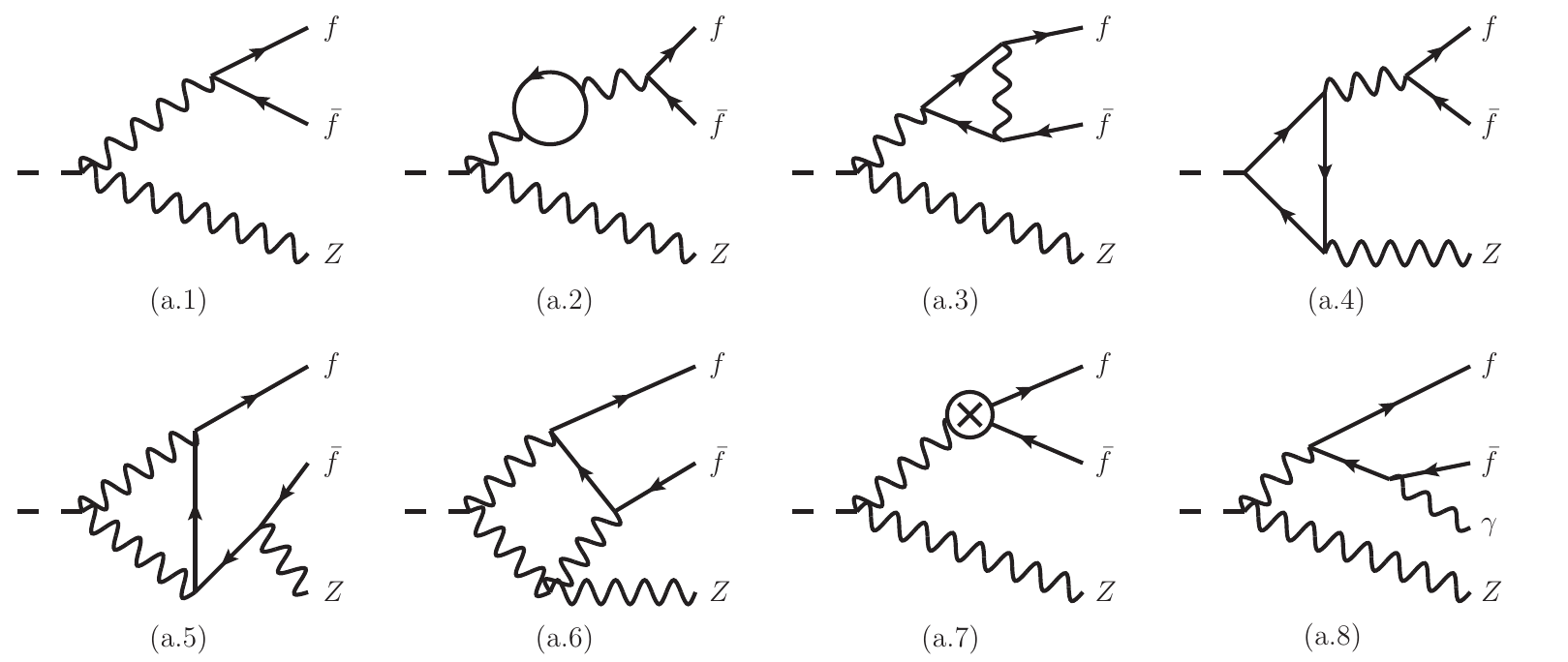}}\\
\subfigure[]{
\includegraphics[scale=0.6]{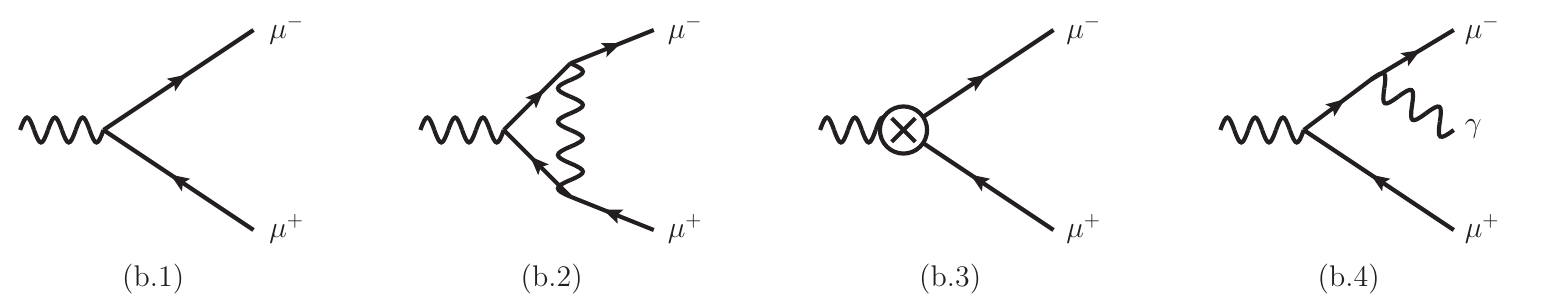}}
\caption{Typical LO and NLO Feynman diagrams for the processes (a) $H\to f\bar{f}Z$, (b) $Z\to \mu^-\mu^+$.}
\label{fig_Feyn}
\end{figure}

In this section, we elucidate some technical details in the calculation of $\Gamma^{H\to ffZ}_\lambda$ and $\Gamma^{Z\to \mu\mu}_\lambda$.
The typical LO and NLO Feynman diagrams are shown in Fig. \ref{fig_Feyn}.
The calculation is carried out in the Feynman gauge, with the masses of leptons and the first two generation quarks neglected.
The Mathematica package FeynArts \cite{Hahn:2000kx} is used to generate the Feynman diagrams as well as the amplitudes.
With the aid of FeynCalc \cite{Mertig:1990an,Shtabovenko:2020gxv}, the amplitudes are then expressed in terms of scalar integrals.
The package LoopTools \cite{Hahn:1998yk} is employed to calculate the scalar integrals numerically.
The numerical phase-space integration is performed with the package VEGAS+ \cite{Lepage:2020tgj}.

The LO results are pretty simple.
For the $H\to f\bar{f}Z$ process, we have
\begin{align}
\label{eq_LOres1}
&\frac{{\rm d}^2\Gamma^{H\to ffZ}_\text{T,LO}}{{\rm d}t\;{\rm d}u}=\frac{\alpha^2 }{32\pi s_{\rm w}^4c_{\rm w}^4}\frac{c_1}{m_H^3}\frac{m_Z^2(m_H^2+m_Z^2-t-u)}{(m_H^2-t-u)^2} \frac{2 m_H^2 m_Z^2-t^2-u^2}{4 m_H^2 m_Z^2-(t+u)^2}, \\
%%%%%%%%%
&\frac{{\rm d}^2\Gamma^{H\to ffZ}_\text{L,LO}}{{\rm d}t\;{\rm d}u}=\frac{\alpha^2}{64\pi s_{\rm w}^4c_{\rm w}^4}\frac{c_1}{m_H^3}\;\frac{(2m_Z^2-t-u)^2}{(m_H^2-t-u)^2}\frac{m_H^2m_Z^2-tu}{4m_H^2m_Z^2-(t+u)^2},\\
%%%%%%%%%
\label{eq_LOres3}
&\frac{{\rm d}^2\Gamma^{H\to ffZ}_\text{A,LO}}{{\rm d}t\;{\rm d}u}=\frac{\alpha^2 }{32\pi s_{\rm w}^4c_{\rm w}^4}\frac{c_2}{m_H^3}\frac{m_Z^2(m_H^2+m_Z^2-t-u)}{(m_H^2-t-u)^2}\frac{t-u}{\sqrt{4 m_H^2 m_Z^2-(t+u)^2}},
\end{align}
where $\alpha$ is the electromagnetic coupling constant, $s_{\rm w}$ and $c_{\rm w}$ are the sine and cosine of the Weinberg angle,
$t=(p_1-p_2)^2$ and $u=(p_1-p_3)^2$ are the Mandelstam variables.
The coefficients $c_i$ for different channels are
\begin{align}
c_1=8s_{\rm w}^4-4s_{\rm w}^2+1,\ c_2=4s_{\rm w}^2-1, \quad &{\rm for}\ f=e,\mu ;\\
c_1=1,\ c_2=1, \quad &{\rm for}\ f=\nu_e,\nu_\mu,\nu_\tau ;\\
c_1=\frac{3}{9}(32s_{\rm w}^4-24s_{\rm w}^2+9),\ c_2=8s_{\rm w}^2-3, \quad &{\rm for}\ f=u,s ;\\
c_1=\frac{3}{9}(8s_{\rm w}^4-12s_{\rm w}^2+9),\ c_2=4s_{\rm w}^2-3, \quad &{\rm for}\ f=d.
\end{align}
For the $Z\to \mu^-\mu^+$ process, we have
\begin{align}
&\frac{{\rm d}\Gamma^{Z\to \mu\mu}_\text{T,LO}}{{\rm d cos}\theta^*}=\frac{\alpha }{64 c_{\rm w}^2 s_{\rm w}^2} m_Z(8s_{\rm w}^4-4s_{\rm w}^2+1)(1+{\rm cos}^2\theta^*),\\
%%%%%%
&\frac{{\rm d}\Gamma^{Z\to \mu\mu}_\text{L,LO}}{{\rm d cos}\theta^*}=\frac{\alpha }{32 c_{\rm w}^2 s_{\rm w}^2} m_Z(8s_{\rm w}^4-4s_{\rm w}^2+1)(1-{\rm cos}^2\theta^*),\\
%%%%%%
&\frac{{\rm d}\Gamma^{Z\to \mu\mu}_\text{A,LO}}{{\rm d cos}\theta^*}=\frac{\alpha }{32 c_{\rm w}^2 s_{\rm w}^2} m_Z (4s_{\rm w}^2-1){\rm cos}\theta^*.
\end{align}

We consider the NLO EW corrections to the $H\to f\bar{f}Z$ and $Z\to \mu^+\mu^-$ processes, where $f$ denotes any light fermions;
and the NLO quantum chromodynamics (QCD) corrections to the $H\to q\bar{q}Z$ process, where $q$ denotes any light quarks.
The typical Feynman diagrams for the NLO corrections are shown in Fig. \ref{fig_Feyn} (a.2)--(a.8) and (b.2)--(b.4).
The conventional dimensional regularization with $D=4-2\epsilon$ is adopted to regularize the ultraviolet (UV) and infrared (IR) singularities.
The method proposed in Refs. \cite{Kreimer:1989ke,Korner:1991sx} is used to deal with the $D$-dimensional $\gamma_5$ trace.

According to the power counting rule, the UV singularities arise from the self-energy and triangle diagrams (like Fig. \ref{fig_Feyn}(a.2)--(a.5)(b.2)), and can be removed through renormalization procedure.
Here we use the counter-term approach.
Following the implementation of Ref. \cite{Denner:1991kt}, we recalculate the renormalization constants by setting the light-fermion mass to exactly zero.
For the electromagnetic coupling constant $\alpha$, we use the $G_\mu$ scheme, with the charge renormalization constant redefined as $\delta Z_e^{G_\mu}=\delta Z_e-\Delta r/2$, where $\Delta r$ is the one-loop weak corrections to muon decay \cite{Denner:2019vbn}.
After including the counter-term diagrams (like Fig. \ref{fig_Feyn}(a.7)(b.3)), all UV singularities cancel each other.

The IR singularities originate from soft or collinear massless particles, and are regulated dimensionally as well.
To handle the IR singularities in the real corrections (like Fig. \ref{fig_Feyn}(a.5)(b.4)), the Catani-Seymour dipole subtraction method \cite{Catani:1996vz} is adopted.
Following the procedure of Ref. \cite{Catani:1996vz}, the correction terms can be rewritten as
\begin{equation}
\int _{m+1}\big[{\rm d}\Gamma_{\rm real}-{\rm d}\Gamma_{\rm A}\big]+\int _{m}\bigg[{\rm d}\Gamma_{\rm virtual}+\int_1{\rm d}\Gamma_{\rm A}\bigg],
\label{eq_dipole}
\end{equation}
where ${\rm d}\Gamma_{\rm real}$ and ${\rm d}\Gamma_{\rm virtual}$ denote the real and virtual corrections respectively,
${\rm d}\Gamma_{\rm A}$ is the auxiliary dipole subtraction term which possesses the same pointwise singular behavior as ${\rm d}\Gamma_{\rm real}$.
As ${\rm d}\Gamma_{\rm A}$ acts as a local counter term for ${\rm d}\Gamma_{\rm real}$, the first integral of Eq. (\ref{eq_dipole}) is nonsingular at every points of phase space, and can be evaluated numerically in four dimensions.
On the other hand, the integration of ${\rm d}\Gamma_{\rm A}$ over one-body subspace, i.e. $\int_1{\rm d}\Gamma_{\rm A}$, can be carried out analytically in $D$ dimensions.
After adding up $\int_1{\rm d}\Gamma_{\rm A}$ and ${\rm d}\Gamma_{\rm virtual}$, all IR singularities are eliminated as expected.

\section{Numerical results}
\label{sec_numerical}
The input parameters taken in the numerical calculation go as follows \cite{ParticleDataGroup:2022pth}:
\begin{align}
\label{eq_paraL1}
&G_\mu=1.16638\times 10^{-5}\ \text{GeV}^{-2},\quad\quad \alpha_s(m_Z)=0.1179 \nonumber \\
& m_H=125.25\ \text{GeV},\quad\quad m_Z=91.19\ {\rm GeV},\quad\quad \Gamma_Z=2.495\ {\rm GeV}, \nonumber \\
&m_W=80.38\ \text{GeV},\quad\quad m_t=172.69\ \text{GeV},\quad\quad m_b=4.18\ \text{GeV};\\
\label{eq_paraL2}
&\alpha(0)=1/137.036,\quad\quad \alpha(m_Z^2)=1/127.951, \quad\quad m_e=0.511\ \text{MeV}, \nonumber \\
&m_\mu=105.66\ \text{MeV},\quad\quad m_\tau=1.77686\ \text{GeV},\quad\quad m_u=66\ \text{MeV},\nonumber \\
&m_d=66\ \text{MeV},\quad\quad m_c=1.27\ \text{GeV}, \quad\quad m_s=150\ \text{MeV}.
\end{align}
Here, the weak-boson masses and the $Z$-boson width are defined by their on-shell values\footnote{Alternatively, one may use the one-loop value of the $Z$-boson width to obtain a consistent one-loop description. 
According to Ref. \cite{Bredenstein:2006rh}, the one-loop $Z$-boson width is estimated to be $\Gamma_Z^\text{1-loop}=2.503\ {\rm GeV}$.
Since in our calculation, the $Z$-boson width term is factored out (see Eq. (\ref{eq_mainformu})), the predictions with $\Gamma_Z^\text{1-loop}$ as input can be obtained by multiplying the original predictions with the overall factor $\big[\frac{1}{\Gamma_Z^\text{1-loop}}\text{arctan}(\frac{s_{45}-m_Z^2}{\Gamma_Z^\text{1-loop} m_Z})|^{s_{45}=s_{45}^+}_{s_{45}=s_{45}^-}\big]/\big[\frac{1}{\Gamma_Z}\text{arctan}(\frac{s_{45}-m_Z^2}{\Gamma_Z m_Z})|^{s_{45}=s_{45}^+}_{s_{45}=s_{45}^-}\big]\approx 0.996$.}.
The weak mixing angle is fixed by $c_{\rm w}=m_W/m_Z$, $s_{\rm w}=\sqrt{1-c_{\rm w}^2}$.
The strong coupling constant $\alpha_s(m_Z)$ is used in the calculation of QCD corrections to the $H\to q\bar{q}Z$ process.

Our main calculations are performed under the $G_\mu$ scheme, with the electroweak coupling $\alpha_{G_\mu}=\sqrt{2}G_\mu m_W^2(1-c_{\rm w}^2)/\pi$.
The results under the $\alpha(0)$ and $\alpha(m_Z^2)$ schemes can be obtained by  making the substitutions $\alpha_{G_\mu}\to \alpha(0)(1+\Delta r)$ and $\alpha_{G_\mu}\to \alpha(m_Z^2)(1+\Delta r-\Delta\alpha(m_Z^2))$. 
The details about the quantities $\Delta r$ and  $\Delta\alpha(m_Z^2)$ can be found in Ref. \cite{Denner:2019vbn}.
By default, we take the results under the $G_\mu$ scheme as the central values, the results under the $\alpha(0)$ and $\alpha(m_Z^2)$ schemes minus the central values as the theoretical uncertainties.
The additional parameters used in the error estimation are listed in Eq. (\ref{eq_paraL2})\footnote{Here, the masses of the $u$, $d$, $s$ quarks are taken from Ref. \cite{Dittmaier:2009cr}, other parameters are taken from the Particle Data Group’s Review of Particle Physics \cite{ParticleDataGroup:2022pth}. This implementation reproduce the hadronic contribution to the photonic vacuum polarization, which takes the value $\Delta^{(5)}_\text{had}(m_Z^2)=0.0276$ \cite{ParticleDataGroup:2022pth}.}.

In the following subsections, we first present the decay widths of the $Z\to \mu^-\mu^+$ and $H\to f\bar{f}Z$ processes separately, and then substitute them into Eq. (\ref{eq_mainformu}) to obtain the final predictions.

\subsection{$Z\to \mu^-\mu^+$}
At the NLO, the distribution $\text{d}\Gamma^{Z\to \mu\mu}_\lambda/\text{dcos}\theta^*$ depends on how we treat the additional photon in the real emission process.
Experimentally, both collinear and noncollinear photons can be identified in the reconstruction of $Z$ boson \cite{ATLAS:2012ana,ATLAS:2012vua}.
For this sake, we define $\theta^*$ of the real emission process as the polar angle of $\mu^+$, in the rest frame of $(\mu^-\mu^+\gamma)$ system, with respect to the $(\mu^-\mu^+\gamma)$ flight direction in the Higgs rest frame.
In such a case, the distribution $\text{d}\Gamma^{Z\to \mu\mu}_\lambda/\text{dcos}\theta^*$ up to NLO can be fitted into the form
\begin{align}
\label{eq_Zf1}
&\frac{1}{\Gamma_Z}\frac{\text{d}\Gamma^{Z\to \mu\mu}_\text{T}}{\text{d}\cos\theta^*}=(c_1+c_2\cos^2\theta^*)\times 10^{-2},\\
%%%%%%%
\label{eq_Zf2}
&\frac{1}{\Gamma_Z}\frac{\text{d}\Gamma^{Z\to \mu\mu}_\text{L}}{\text{d}\cos\theta^*}=(c_3-c_4\cos^2\theta^*)\times 10^{-2},\\
%%%%%%%
\label{eq_Zf3}
&\frac{1}{\Gamma_Z}\frac{\text{d}\Gamma^{Z\to \mu\mu}_\text{A}}{\text{d}\cos\theta^*}=(-c_5\cos\theta^*)\times 10^{-2}.
\end{align}
The LO and NLO values of $c_i$ are presented in Table \ref{tab_ci}, wherein, the central values refer to the results under the $G_\mu$ scheme, the supers- and subscripts corresponding to the results under the $\alpha(m_Z^2)$ and $\alpha(0)$ schemes, respectively.
It can be seen that the EW corrections are generally of order or less than $0.1\%$, except for $c_5$, where the correction reaches $34\%$.
As we will see in the next subsection, the asymmetry width $\Gamma_\text{A}^{H\to ffZ}$ vanishes up to NLO, which indicates that $c_5$ has no impact on the final results.
We also find that the EW corrections greatly suppress the scheme dependence and enhance the prediction reliability.

\begin{table}
   \centering
%  \fontsize{6.5}{8}\selectfont
  \caption{The LO and NLO predictions to the $c_i$ of Eqs. (\ref{eq_Zf1})-(\ref{eq_Zf3}).}
  \label{tab_ci}
    \begin{tabular}{|p{1.2cm}<{\centering}|p{2.2cm}<{\centering}| p{2.2cm}<{\centering}|p{2.2cm}<{\centering}|p{2.2cm}<{\centering}|p{2.2cm}<{\centering}|}
    \hline
    & $c_1$ & $c_2$ & $c_3$ & $c_4$ & $c_5$ \cr\hline 
     LO & $1.261^{+0.042}_{-0.045}$ & $1.261^{+0.042}_{-0.045}$ & $2.522^{+0.083}_{-0.090}$ & $2.522^{+0.083}_{-0.090}$ & $0.538^{+0.018}_{-0.019}$ \cr\hline
      NLO & $1.262^{+0.001}_{-0.008}$ & $1.260^{+0.001}_{-0.008}$ & $2.522^{+0.002}_{-0.016}$ & $2.520^{+0.002}_{-0.016}$ & $0.355^{-0.012}_{+0.009}$ \cr\hline
      \end{tabular}
\end{table}

\subsection{$H\to f\bar{f}Z$}
In the calculation of $\Gamma^{H\to ffZ}_\lambda$, we employ three different kinematic cuts on the invariant mass of the final state products except the $Z$ boson:
 no cut, $M_{ff(\gamma)}>6$ GeV, 6 GeV$<M_{ff(\gamma)}<29$ GeV.
The second cut excludes the soft and collinear fermion pair, which mainly originates from the $H\to \gamma Z$ process.
The third cut sets an upper bound to $M_{ff(\gamma)}$, which can reduce the theoretical error caused by the on-shell approximation of the phase-space integral (see Eq. (\ref{eq_dI3approx})).

The LO and NLO decay widths of various channels of $H\to f\bar{f}Z$ are shown in Table \ref{tab_HffZnum}.
As in Table \ref{tab_ci}, the central values are calculated under the $G_\mu$ scheme, the super- and subscripts lead to the results under the $\alpha(m_Z^2)$ and $\alpha(0)$ schemes, respectively.
Also, after including the NLO corrections, the scheme dependence is reduced significantly.
It should be remarked that the asymmetry decay width $\Gamma^{H\to ffZ}_\text{A}$ vanishes up to NLO.
We verified analytically that the difference $ \big|\mathcal{M}_{++}^{H\to ffZ(
\gamma)}\big|^2-\big|\mathcal{M}_{--}^{H\to ffZ(\gamma)}\big|^2$ is asymmetry under the exchange $p_2 \leftrightarrow p_3$, which results in $\Gamma^{H\to ffZ}_\text{A}=0$ after the phase-space integration.
For the transverse and longitudinal widths, the NLO corrections are $1\%$--$4\%$ of the LO contributions, which exhibits good convergence behavior of the perturbation theory.
We also present the longitudinal fraction $f_\text{L}=\Gamma^{H\to ffZ}_\text{L}/(\Gamma^{H\to ffZ}_\text{T}+\Gamma^{H\to ffZ}_\text{L})$ in Table \ref{tab_HffZnum}.
It can be seen that the NLO corrections to $f_\text{L}$s are insignificant, which are less than $1\%$ for all channels.
Note, the difference of $f_\text{L}$s under different $\alpha$ schemes are hardly visible.

\begin{table}
   \centering
%  \fontsize{6.5}{8}\selectfont
  \caption{The LO and NLO decay widths of various channels of $H\to f\bar{f}Z$. The longitudinal fraction is defined as $f_\text{L}=\Gamma^{H\to ffZ}_\text{L}/(\Gamma^{H\to ffZ}_\text{T}+\Gamma^{H\to ffZ}_\text{L})$. }
  \label{tab_HffZnum}
    \begin{tabular}{|p{2.2cm}<{\centering}| p{2.3cm}<{\centering}|p{1.8cm}<{\centering}|p{1.8cm}<{\centering}|p{1.8cm}<{\centering}|p{1.8cm}<{\centering}|p{1.8cm}<{\centering}|p{1.8cm}<{\centering}|}
    \hline
       \multirow{2}{*}{channel} & \multirow{2}{*}{polarization} &  \multicolumn{2}{c|}{no cut} & \multicolumn{2}{c|}{$M_{ff(\gamma)}>6$} &  \multicolumn{2}{c|}{$6<M_{ff(\gamma)}< 29$} \cr \cline{3-8} 
     & &  LO & NLO & LO & NLO & LO & NLO \cr \hline
 %%%%%%%%%%%     
      \multirow{3}{*}{\makecell{$H\to \ell^-\ell^+ Z$,\\ $\ell=e,\mu$}} & $\Gamma_\text{T}^{H\to \ell\ell Z}$(keV) & $1.218_{-0.085}^{+0.082}$ & $1.233_{-0.018}^{+0.002}$ & $1.216_{-0.085}^{+0.081}$ & $1.232_{-0.018}^{+0.002}$ & $0.820_{-0.057}^{+0.055}$ & $0.827_{-0.012}^{+0.001}$ \cr \cline{2-8}
    &$\Gamma_\text{L}^{H\to \ell\ell Z}$(keV) & $1.786_{-0.125}^{+0.120}$ & $1.813_{-0.028}^{+0.004}$ & $1.702_{-0.119}^{+0.114}$ & $1.728_{-0.026}^{+0.004}$ & $1.450_{-0.101}^{+0.097}$ & $1.472_{-0.022}^{+0.003}$ \cr \cline{2-8}
      &$f_\text{L}$ & 0.594 & 0.595 & 0.583 & 0.584 & 0.639 & 0.640 \cr\hline\hline
 %%%%%%%%%%%%
      \multirow{3}{*}{\makecell{$H\to \nu_\ell \bar{\nu}_\ell Z$,\\ $\ell=e,\mu,\tau$}} & $\Gamma_\text{T}^{H\to \nu\nu Z}$(keV) & $2.408_{-0.168}^{+0.161}$ & $2.475_{-0.040}^{+0.008}$ & $2.404_{-0.168}^{+0.161}$ & $2.471_{-0.040}^{+0.008}$ & $1.621_{-0.113}^{+0.109}$ & $1.666_{-0.027}^{+0.006}$ \cr \cline{2-8}
       &$\Gamma_\text{L}^{H\to \nu\nu Z}$(keV) & $3.530_{-0.246}^{+0.236}$ & $3.600_{-0.056}^{+0.009}$ & $3.364_{-0.235}^{+0.225}$ & $3.431_{-0.054}^{+0.009}$ & $2.867_{-0.200}^{+0.192}$ & $2.921_{-0.045}^{+0.007}$ \cr \cline{2-8}
      &$f_\text{L}$ & 0.594 & 0.592 & 0.583 & 0.581 & 0.639 & 0.637 \cr\hline\hline
 %%%%%%%%%%%%
      \multirow{3}{*}{\makecell{$H\to q \bar{q} Z$, \\ $q=u$}} & $\Gamma_\text{T}^{H\to qq Z}$(keV) & $4.206_{-0.293}^{+0.282}$ & $4.285_{-0.061}^{+0.005}$ & $4.199_{-0.293}^{+0.281}$ & $4.286_{-0.062}^{+0.006}$ & $2.831_{-0.198}^{+0.190}$ & $2.863_{-0.039}^{+0.001}$ \cr \cline{2-8}
      &$\Gamma_\text{L}^{H\to qq Z}$(keV) & $6.165_{-0.430}^{+0.413}$ & $6.377_{-0.100}^{+0.017}$ & $5.875_{-0.410}^{+0.394}$ & $6.079_{-0.095}^{+0.017}$ & $5.006_{-0.349}^{+0.335}$ & $5.175_{-0.081}^{+0.014}$ \cr \cline{2-8}
      &$f_\text{L}$ & 0.594 & 0.598 & 0.583 & 0.586 & 0.639 & 0.644 \cr\hline\hline
 %%%%%%%%%%%%
      \multirow{3}{*}{\makecell{$H\to q \bar{q} Z$, \\ $q=d,s$}} & $\Gamma_\text{T}^{H\to qq Z}$(keV) & $5.396_{-0.376}^{+0.362}$ & $5.572_{-0.086}^{+0.014}$ & $5.387_{-0.376}^{+0.361}$ & $5.570_{-0.087}^{+0.015}$ & $3.633_{-0.254}^{+0.243}$ & $3.732_{-0.056}^{+0.008}$ \cr \cline{2-8}
       &$\Gamma_\text{L}^{H\to qq Z}$(keV) & $7.910_{-0.552}^{+0.530}$ & $8.229_{-0.133}^{+0.027}$ & $7.537_{-0.526}^{+0.505}$ & $7.845_{-0.127}^{+0.026}$ & $6.423_{-0.448}^{+0.430}$ & $6.678_{-0.107}^{+0.022}$ \cr \cline{2-8}
      &$f_\text{L}$ & 0.594 & 0.596 & 0.583 & 0.585 & 0.639 & 0.641 \cr\hline\hline
      \multicolumn{8}{|c|}{$\Gamma^{H\to ffZ}_\text{A}=0$ for all channels.} \cr \hline
      \end{tabular}
\end{table}

\subsection{$H\to f\bar{f}Z\to f\bar{f}\mu^-\mu^+$}
With the results presented in the above two subsections, the predictions to $H\to f\bar{f}Z\to f\bar{f}\mu^-\mu^+$  can be obtained by employing Eq. (\ref{eq_mainformu}).
To estimate the contributions beyond the leading pole approximation, we introduce the quantity
\begin{equation}
\Delta_\text{BLP}=\Gamma^\text{full}_\text{LO}-\Gamma_\text{LO}^\text{SPA}, 
\end{equation}
where $\Gamma^\text{full}_\text{LO}$ is the full LO decay width which calculated in the complex-mass scheme \cite{Denner:1999gp,Denner:2006ic}.

\begin{figure}[h!]
\centering
\subfigure[]{
\includegraphics[width=0.48\textwidth]{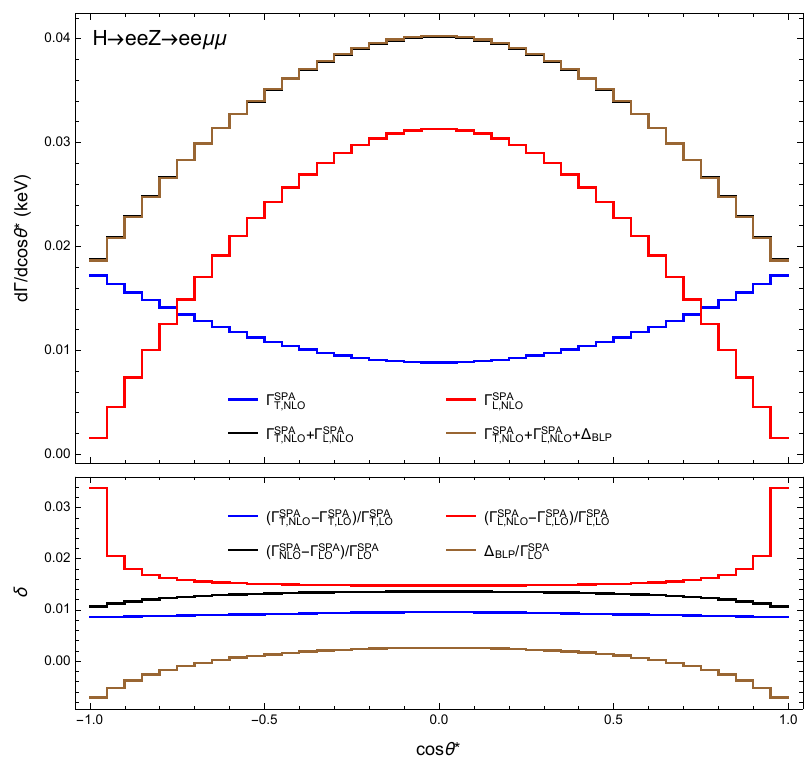}}
\subfigure[]{
\includegraphics[width=0.48\textwidth]{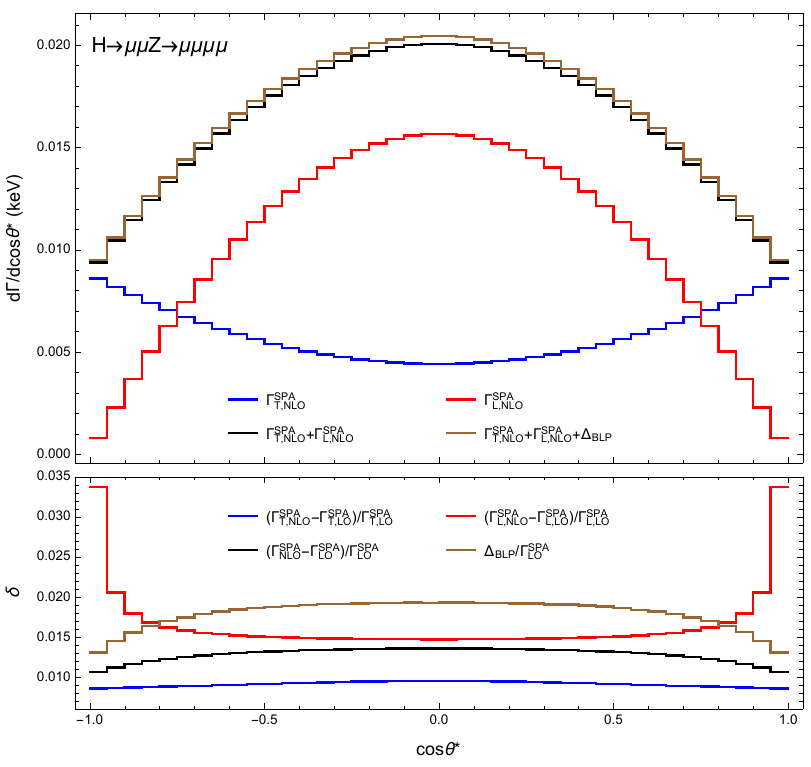}}
\subfigure[]{
\includegraphics[width=0.48\textwidth]{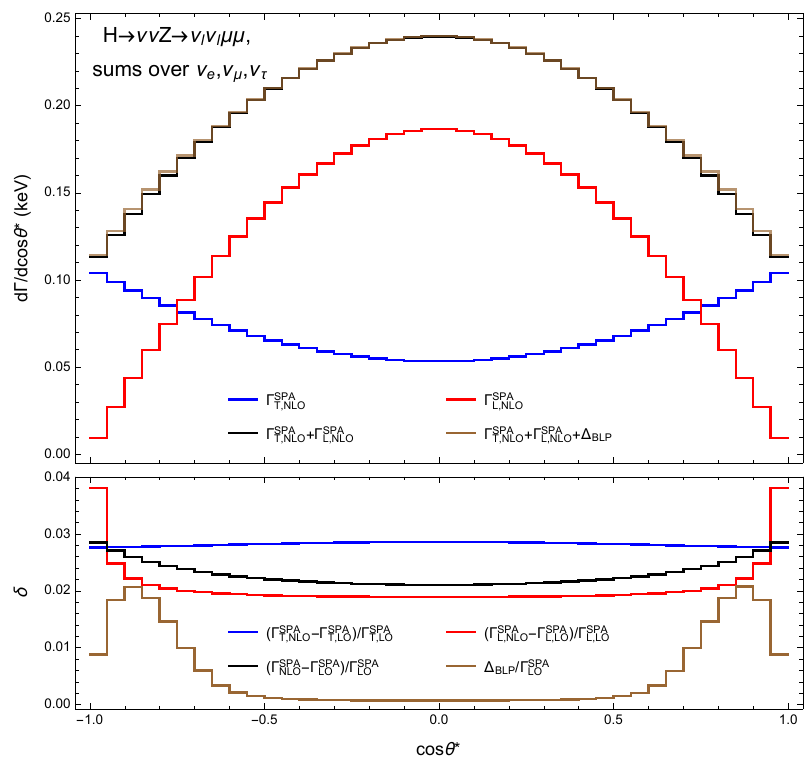}}
\subfigure[]{
\includegraphics[width=0.48\textwidth]{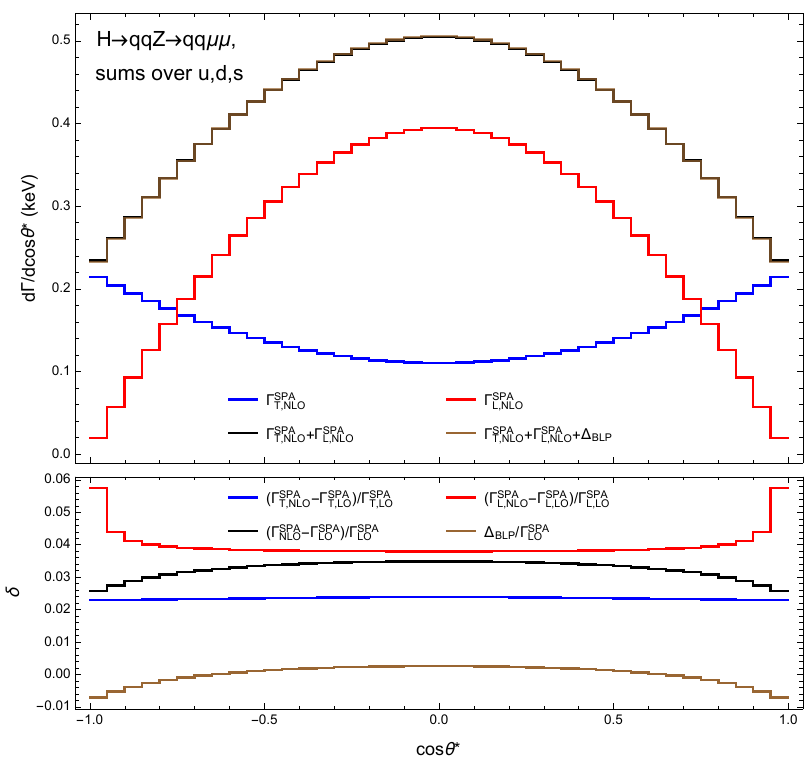}}
\caption{Differential decay widths with respect to $\text{cos}\theta^*$ for the $H\to f\bar{f}Z\to f\bar{f}\mu^-\mu^+$ process. (a) $f=e$; (b) $f=\mu$; (c) $f$ sums over $\nu_e,\nu_\mu,\nu_\tau$; (d) $f$ sums over $u,d,s$.}
\label{fig_resdis}
\end{figure}

The predictions under the $G_\mu$ scheme are presented in Fig. \ref{fig_resdis}, 
wherein the kinematic cuts 6 GeV$<M_{ff(\gamma)}<29$ GeV, and $s_{45}^\pm=(m_Z\pm 5\;\text{GeV})^2$ are employed.

The distributions with respect to $\text{cos}\theta^*$ for the $H\to e^-e^+Z\to e^-e^+\mu^-\mu^+$ process are shown in Fig. \ref{fig_resdis}(a).
We present the differential decay widths in the upper panel.
It can be seen that, the longitudinal distribution (the red line in the upper panel) features a maximum at  $\text{cos}\theta^*=0$, and a minimum at $|\text{cos}\theta^*|=1$.
While the transverse distribution (the blue line in the upper panel) exhibits opposite  convexity.
The noticeable difference in shape between these two distributions indicates that the distribution in $\text{cos}\theta^*$ is well suited for polarization discrimination.
The impact of the NLO corrections and the effect beyond the leading pole approximation are demonstrated in the lower panel.
Due to the absence of the non-factorizable diagram at the LO, the leading pole approximation works well, especially at the region $|\text{cos}\theta^*|<0.7$, where the relative $\Delta_\text{BLP}$-correction (the brown line in the lower panel) is negligible.
One may notice that the red line in the lower panel rises as $|\text{cos}\theta^*|$ approaching to $1$.
That is understandable, since the LO longitudinal distribution tends to vanish at this region.

The distributions for the $H\to \mu^-\mu^+Z\to \mu^-\mu^+\mu^-\mu^+$ process are shown in Fig. \ref{fig_resdis}(b).
In comparison with the $H\to e^-e^+Z\to e^-e^+\mu^-\mu^+$ process, the decay widths here gets a factor $1/4$ for identical particles in the final state, and a factor $2$ for possible permutation of the $\mu^-\mu^+$ pair which near the $Z$ pole.
Hence the SPA results here are exactly $1/2$ of those in the Fig. \ref{fig_resdis}(a).
One the other hand, since there is one non-factorizable diagram at the LO, the relative $\Delta_\text{BLP}$-correction (the brown line in the lower panel) is even more significant than the NLO EW corrections.
Fortunately, the relative $\Delta_\text{BLP}$-correction is flat at the region $|\text{cos}\theta^*|<0.7$, which approximately equivalent to an overall factor on the SPA distribution.
For this sake, the non-negligible $\Delta_\text{BLP}$-correction will not spoil the polarization discrimination, if an appropriate angular cut is applied.

The distributions for the $H\to \nu_\ell \bar{\nu}_\ell Z\to \nu_\ell \bar{\nu}_\ell \mu^-\mu^+$ process, where $\ell$ runs over $e,\mu,\tau$, are shown in Fig. \ref{fig_resdis}(c).
Since the angular dependence of the SPA differential decay widths are determined by the $Z\to \mu^-\mu^+$ process (Eqs. (\ref{eq_Zf1})(\ref{eq_Zf2})), the curves in the upper panel are similar in shape as those presented in Fig. \ref{fig_resdis}(a)(b).
It is noticeable that the relative $\Delta_\text{BLP}$-correction (the brown line in the lower panel) exhibits peaks around $|\text{cos}\theta^*|=0.85$, which can be attributed to the $W$-induced diagram of the $H\to \nu_\mu \bar{\nu}_\mu \mu^-\mu^+$ process.
In order to suppress the contribution from the $\Delta_\text{BLP}$-correction, one may introduce an angular cut, like $|\text{cos}\theta^*|<0.7$.

The distributions for the $H\to q \bar{q} Z\to q \bar{q} \mu^+\mu^-$ process, where $q$ sums over $u,d,s$, are shown in Fig. \ref{fig_resdis}(d).
Since both QCD and EW corrections are involved in this process, the relative NLO corrections are more significant than those of other channels.
Also, the relative $\Delta_\text{BLP}$-correction is negligible at the region $|\text{cos}\theta^*|<0.7$.

\section{Summary}
Testing various properties of Higgs boson is one of the key issues in current high energy experiment. 
To this aim, we investigate the Higgs boson decay process $H\to f\bar{f}Z\to f\bar{f}\mu^-\mu^+$, where $f$ denotes any light fermions, at the NLO accuracy.
By combining the SPA and the phase-space decomposition technique, we show that the decay width of $H\to f\bar{f}Z\to f\bar{f}\mu^-\mu^+$ can be factorized into the width of $H\to f\bar{f}Z$, the width of $Z\to \mu^-\mu^+$, and an overall factor which mimic the finite width effect of the intermediate $Z$ boson.
The phenomenological analysis is carried out focusing on the polar angular distribution of the antimuon in the rest frame of the $Z$ boson.
Numerical results show that, by taking appropriate cut, the non-resonant background is negligible, and the NLO corrections are $1\%$--$4\%$ of the LO contributions.
To tame the theoretical uncertainty, we adopt three different electroweak coupling  schemes, and find that the NLO corrections can reduce the scheme dependence as expected.
To investigate the polarization effects of the intermediate $Z$ boson, the contributions of different polarization modes are differentiated. 
We find that the polarization fractions (polarized widths over the unpolarized one) are about $0.4$ for the transverse modes, $0.6$ for the longitudinal ones.

%%%%%%%%%%%%%%%%%%%%%%%%%%%%%%%%%%%%%%%%%%%%%%%%%%%%%%%%%%%%%%%%%%%%%

\vspace{1cm} {\bf ACKNOWLEDGMENTS}

This work was supported in part by the National Key Research and Development Program of China under Contract 2020YFA0406400,
and the National Natural Science Foundation of China (NSFC) under the Grants 12205061, 12047553, 12235008 and 12475087.
%%%%%%%%%%%%%%%%%%%%%%%%%%%%%%%%%%%%%%%%%%%%%%%%%%%%%%%%%%%%%%%%%%%%

\vspace{0.7cm}

\section*{APPENDIX: PROOF OF EQ. (\ref{eq_intfis0})}
In this appendix, we present the proof of Eq. (\ref{eq_intfis0}) at the NLO accuracy.
We first consider the two-body decay process $Z(p_{45})\to \mu^-(p_4)+\mu^+(p_5)$, from which the LO and the virtual correction terms can be constructed. The external momenta are assigned as 
\begin{align}
&p_{45}=m_Z\{1,0,0,0\},\quad p_{4}=\frac{m_Z}{2}\{1,-\sin\theta^*\cos\phi^*,-\sin\theta^*\sin\phi^*,-\cos\theta^* \},\nonumber \\
&p_5=p_{45}-p_4,
\label{eq_ap_mom2}
\end{align}
and the polarization vectors of the $Z$ boson are chosen as 
\begin{align}
\varepsilon_+=\frac{1}{\sqrt{2}}\{0,1,-i,0\},\quad \varepsilon_-=\frac{1}{\sqrt{2}}\{0,-1,-i,0\},\quad \varepsilon_0=\{0,0,0,1\}.
\label{eq_ap_pol}
\end{align}
Considering the constraint $p_{45}\cdot \varepsilon_\lambda=0$, the Lorentz structure of the squared amplitude is
\begin{align}
 \big[\mathcal{M}^{Z\to \mu\mu}_{\mu} \varepsilon^{\mu}_{\lambda}\big]\big[\mathcal{M}^{Z\to \mu\mu}_{\nu} \varepsilon^{\nu}_{\lambda^\prime}\big]^*=\varepsilon^{\mu}_{\lambda}\varepsilon^{* \nu}_{\lambda^\prime}\left[ 
 A(m_Z)(p_4-p_5)^\mu (p_4-p_5)^\nu+B(m_Z) \epsilon^{p_4p_5\mu\nu}\right].
 \label{eq_ap_lstr}
\end{align}
Substituting Eqs. (\ref{eq_ap_mom2})(\ref{eq_ap_pol}) into Eq. (\ref{eq_ap_lstr}), it is straightforward to show that 
\begin{equation}
\int^{2\pi}_0 {\rm d}\phi^* \big[\mathcal{M}^{Z\to \mu\mu}_{\mu} \varepsilon^{\mu}_{\lambda}\big]\big[\mathcal{M}^{Z\to \mu\mu}_{\nu} \varepsilon^{\nu}_{\lambda^\prime}\big]^*
= 0,\; \text{for}\; \lambda\ne \lambda^\prime.
\end{equation}

For the real corrections, the three-body decay process $Z(p_{45})\to \mu^-(p_4)+\mu^+(p_5)+\gamma(p_\gamma)$ is involved.
Since the additional photon can be identified in the reconstruction of the $Z$ boson \cite{ATLAS:2012ana,ATLAS:2012vua}, the momenta of final-state particles can be defined without any lepton-photon recombination.
Without loss of generality, we choose the following momentum configuration: 
\begin{align}
&p_{45}=m_Z\{1,0,0,0\},\quad p_4=E_4\{1,\sin\theta^\prime \cos(\phi^*+\alpha),\sin\theta^\prime\sin(\phi^*+\eta),\cos\theta^\prime\}, \nonumber \\
& p_5=E_5\{1,\sin\theta^* \cos\phi^*,\sin\theta^*\sin\phi^*,\cos\theta^*\},\quad p_\gamma=p_{45}-p_4-p_5.
\label{eq_ap_realmom}
\end{align}
Here, $E_5$ can be eliminated by using $p_\gamma^2=0$:
\begin{equation}
E_5= \frac{m_Z (m_Z-2 E_4)}{2(E_4\sin \theta^\prime \sin \theta^* \cos\eta+E_4\cos \theta^\prime \cos \theta^* -E_4+m_Z)}.
\end{equation}
With the polarization vectors defined in Eq. (\ref{eq_ap_pol}), we have
\begin{align}
\label{eq_ap_realin1}
&\big[\mathcal{M}^{Z\to \mu\mu\gamma}_{\mu} \varepsilon^{\mu}_+\big]\big[\mathcal{M}^{Z\to \mu\mu\gamma}_{\nu} \varepsilon^{\nu}_-\big]^*=\frac{32(\pi\alpha)^2(8 s_\text{w}^4-4 s_\text{w}^2+1)}{c_\text{w}^2 s_\text{w}^2}\frac{m_Z^2}{ t u}(p_{4,-}^2+p_{5,-}^2),\\
\label{eq_ap_realin2}
&\big[\mathcal{M}^{Z\to \mu\mu\gamma}_{\mu} \varepsilon^{\mu}_\pm \big]\big[\mathcal{M}^{Z\to \mu\mu\gamma}_{\nu} \varepsilon^{\nu}_0\big]^*=\frac{4(\pi \alpha)^2 (4s_\text{w}^2-1)}{c_\text{w}^2 s_\text{w}^2}\frac{m^{}_Z}{t^2u^2}\bigg\{ \bigg[(6t^2-2 t u-u^2)(E_4^2+p_{4,z}^2)\nonumber \\
&\quad\quad+2(4t^2-2t u+u^2)p^{}_{4,z}p^{}_{5,z}+2m_Z^2u(2t+u)-u(2t+u)^2-2m^{}_Z(t^2+u^2)E_4\nonumber\\
&\quad\quad+(t+u)(6t^2-4t u+u^2)\frac{E_4}{m_Z}\pm 8m^{}_Z\frac{1-4s_\text{w}^2+8s_\text{w}^4}{1-4s_\text{w}^2}t u  p^{}_{5,z}\bigg]p^{}_{5,\mp}\nonumber \\
&\quad\quad+\bigg[-(4t^2-2 t u+u^2)(E_5^2+p_{5,z}^2)-2(6t^2-2t u-u^2)p^{}_{4,z}p^{}_{5,z}-2m_Z^2t(t+2u)\nonumber \\
&\quad\quad+t(6t^2+4tu-u^2)+2m^{}_Z(t^2+u^2)E_5-(t+u)(6t^2-4t u+u^2)\frac{E_5}{m_Z}\nonumber\\
&\quad\quad\pm 8m^{}_Z\frac{1-4s_\text{w}^2+8s_\text{w}^4}{1-4s_\text{w}^2}t u  p^{}_{4,z}\bigg]p^{}_{4,\mp}-2(6t^2-2tu-u^2)p_{4,\mp}^2p_{5,\pm}^{}\nonumber\\
&\quad\quad+2(4t^2-2tu+u^2)p_{5,\mp}^2p_{4,\pm}^{} \bigg\}.
\end{align}
Here, for brevity, we introduce the notation $p^{}_{i,\pm}=(p^{}_{i,x}\pm i p^{}_{i,y})/\sqrt{2}$.
We also introduce the Mandelstam variables $t=(p_5+p_\gamma)^2$, $u=(p_4+p_\gamma)^2$.
Substituting Eq. (\ref{eq_ap_realmom}) into Eqs. (\ref{eq_ap_realin1})(\ref{eq_ap_realin2}), it is straightforward to show that 
\begin{equation}
\int^{2\pi}_0 {\rm d}\phi^* \big[\mathcal{M}^{Z\to \mu\mu\gamma}_{\mu} \varepsilon^{\mu}_{\lambda}\big]\big[\mathcal{M}^{Z\to \mu\mu\gamma}_{\nu} \varepsilon^{\nu}_{\lambda^\prime}\big]^*
= 0,\; \text{for}\; \lambda\ne \lambda^\prime.
\end{equation}

\end{document}